\documentclass[conference]{IEEEtran}
\makeatletter
\def\ps@headings{%
\def\@oddhead{\mbox{}\scriptsize\rightmark \hfil \thepage}%
\def\@evenhead{\scriptsize\thepage \hfil \leftmark\mbox{}}%
\def\@oddfoot{}%
\def\@evenfoot{}}
\makeatother
\usepackage{cite}
\usepackage{amsmath,amssymb,amsfonts}
\usepackage{algorithmic}
\usepackage{algorithm}
\usepackage{graphicx}
\usepackage{textcomp}
\usepackage{xcolor}
\usepackage{multirow}
\usepackage{subfigure}
\usepackage{tabularx}
\usepackage{makecell}
\usepackage{enumerate}
\usepackage{booktabs}
\usepackage{tabularray}
\usepackage{adjustbox}
\usepackage{bbding}
\usepackage{tikz}\usepackage{tikz}
\newcommand*\emptycirc[1][1ex]{\tikz\draw (0,0) circle (#1);} 
\newcommand*\halfcirc[1][1ex]{%
	\begin{tikzpicture}
	\draw[fill] (0,0)-- (90:#1) arc (90:270:#1) -- cycle ;
	\draw (0,0) circle (#1);
	\end{tikzpicture}}
\newcommand*\fullcirc[1][1ex]{\tikz\fill (0,0) circle (#1);} 
\usepackage{threeparttable}
\usepackage{booktabs}

\usepackage{geometry}
\def\BibTeX{{\rm B\kern-.05em{\sc i\kern-.025em b}\kern-.08em
   T\kern-.1667em\lower.7ex\hbox{E}\kern-.125emX}}
\begin{document}

\title{Collusion-Driven Impersonation Attack on Channel-Resistant RF Fingerprinting}

\author{
\IEEEauthorblockN{Zhou Xu}
\IEEEauthorblockA{\textit{\shortstack{School of Cyber Science \\and Engineering}} \\
\textit{Southeast University}\\
Nanjing, China \\
xu\_zhou@seu.edu.cn}
\and
\IEEEauthorblockN{Guyue Li}
\IEEEauthorblockA{\textit{\shortstack{School of Cyber Science \\and Engineering}} \\
\textit{Southeast University} \\
Nanjing, China \\
guyuelee@seu.edu.cn\\
(Corresponding author) }
\and
\IEEEauthorblockN{Zhe Peng}
\IEEEauthorblockA{\textit{\shortstack{Dept. of Industrial \\and Systems Engineering}} \\
\textit{\shortstack{The Hong Kong \\Polytechnic University}}\\
Hong Kong, China \\
jeffrey-zhe.peng@polyu.edu.hk}
\and
\IEEEauthorblockN{Aiqun Hu}
\IEEEauthorblockA{\textit{\shortstack{National Mobile Communications \\Research
 Laboratory}} \\
\textit{Southeast University}\\
Nanjing, China \\
aqhu@seu.edu.cn}
}

\maketitle

\begin{abstract}
Radio frequency fingerprint (RFF) is a promising device identification technology, with recent research shifting from robustness to security due to growing concerns over vulnerabilities. To date, while the security of RFF against basic spoofing such as MAC address tampering has been validated, its resilience to advanced mimicry remains unknown. 
To address this gap, we propose a collusion-driven impersonation attack that achieves RF-level mimicry, successfully breaking RFF identification systems across diverse environments. Specifically, the attacker synchronizes with a colluding receiver to match the centralized logarithmic power spectrum (CLPS) of the legitimate transmitter; once the colluder deems the CLPS identical, the victim receiver will also accept the forged fingerprint, completing RF-level spoofing. 
Given that the distribution of CLPS features is relatively concentrated and has a clear underlying structure, we design a spoofed signal generation network that integrates a variational autoencoder (VAE) with a multi-objective loss function to enhance the similarity and deceptive capability of the generated samples. We carry out extensive simulations, validating cross-channel attacks in environments that incorporate standard channel variations including additive white Gaussian noise (AWGN), multipath fading, and Doppler shift. The results indicate that the proposed attack scheme essentially maintains a success rate of over 95\% under different channel conditions, revealing the effectiveness of this attack.
\end{abstract}


\section{Introduction}
\label{sec1}
With the rapid proliferation of Internet of Things (IoT) and wireless communication technologies, device authentication becomes essential to ensure network security
~\cite{10879488}. Radio frequency fingerprinting (RFF), which exploits the inherent hardware impairments of radio transmitters for device identification, has emerged as a promising non-cryptographic physical-layer solution~\cite{soltanieh2020review}. Compared with traditional cryptographic methods, RFF-based techniques avoid key management and require minimal protocol overhead, making them suitable for resource-constrained IoT environments. Over the past decade, the research agenda around RFF has undergone a notable pivot: early efforts concentrated almost exclusively on improving robustness under benign channel variations like multipath fading, Doppler spread, and thermal noise. These studies validated that, once trained on sufficiently diverse channel conditions, RFF classifiers could maintain high accuracy without requiring frequent re-calibration~\cite{yu2019robust}.

\begin{figure}[t]
\centering
\includegraphics[width=\linewidth]{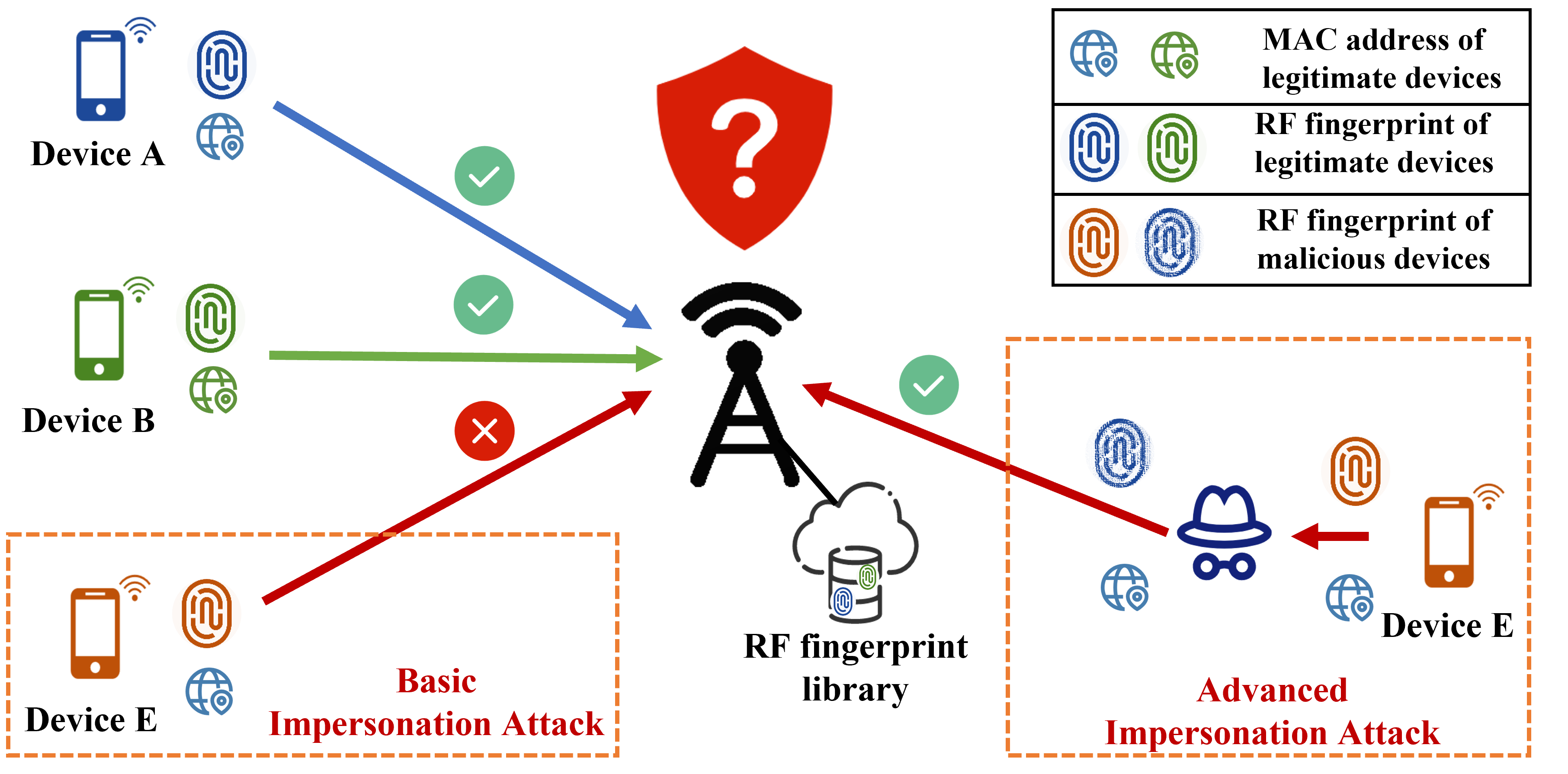}
\caption{Beyond Basic Impersonation: the missed threat of Advanced RF mimicry.}
\label{concept}
\end{figure}

Recently, the community's focus has decisively pivoted towards RFF vulnerabilities which stem from two sources: the neural network-based recognition system and the RF features it relies on. Driven by deep learning’s ability to extract effective features directly from I/Q samples, researchers have widely adopted neural networks for RFF classification, yet this shift simultaneously exposes the system to evasion, backdoor, and impersonation attacks akin to those plaguing computer vision and NLP~\cite{ma2023white, liu2023robust, huang2023hidden, danev2010attacks, liang2023styless, merchant2019securing, karunaratne2021penetrating, dai2024advdiff}. Yet these threats somewhat remain loosely coupled to RFF: they generally rely on strong assumptions that limit their real-world impact. 
Fig.~\ref{concept} spotlights another blind spot in the security of RF features: while basic impersonation attacks are easily caught, the more sophisticated advanced impersonation attack may slip through unnoticed. 
The prevailing belief is that RF fingerprint offers solid security: an attacker who merely forges a MAC address still exposes a distinct RF fingerprint and is therefore detectable; we call this the Basic Impersonation Attack. Yet attackers may go further and fabricate the RF fingerprint itself to deceive the authentication system, an Advanced Impersonation Attack. 
The feasibility of such an attack is critical for the credibility of RF fingerprint security, yet it remains largely uninvestigated and its potential success is uncertain.

RF-level impersonation is challenging primarily because the attacker's receiver is distinct from the legitimate receiver, rendering precise RFF information inaccessible. 
However, this very limitation can be overcome through a collusion attack. The attacker places an RFF module at the colluder matching the legitimate receiver. It captures RF from both sources, extracts fingerprints, infers the target’s features, and crafts spoofed signals indistinguishable to the receiver. This colluding attack framework can reduce the attacker's reliance on prior knowledge. Inspired by this insight, Xu et al.~\cite{xu2022colluding} inserted a colluder and, under additive white Gaussian noise (AWGN) channels, successfully breached the RFF system to achieve RF-level forgery. However, the method cannot be extended to richer multipath environments. It relies on the premise that if the colluder cannot distinguish the attacker’s RF signature from the legitimate one, the legitimate receiver will likewise fail. Because the colluder and the legitimate receiver occupy different physical locations, this premise seldom holds. 
Fortunately, we show this limitation is surmountable. To work in multipath channels, legitimate receivers must rely on channel-robust RF features that yield identical fingerprints at different locations. For example, methods such as the amplitude-only channel-independent spectrogram~\cite{shen2022towards}, amplitude-phase channel-independent spectrogram~\cite{mohammad2023learning}, and centralized logarithmic power spectrum (CLPS)~\cite{tang2024causal} have demonstrated the ability to suppress channel-induced distortions while preserving unique device characteristics. By equipping the colluder with the same feature extractor, we can produce spoofed signals that remain indistinguishable across varying propagation paths, rendering the attack resilient to multipath effects. 
 
Motivated by these considerations, this paper investigates the feasibility and effectiveness of a collusion-driven impersonation attack towards RFF identification in multipath environments. We analyze the limitations of current attack models under realistic channel assumptions and investigate how attackers can exploit channel-resistant feature extraction methods to improve impersonation success rates. 
By introducing CLPS into the attack model and leveraging a variational autoencoder (VAE) for signal generation, we demonstrate that the collusion-driven impersonation attack can achieve high attack success rates across a wide range of challenging wireless environments.

The main contributions of this paper can be summarized as follows:

\begin{itemize}
\item We propose a collusion-driven impersonation attack strategy. To address the issue with existing impersonation attacks, we construct a new attack model. This model introduces CLPS as the channel-resistant RFF, and with the assistance of a colluder, enables the attacker to generate signals that can effectively mimic the target device's channel-resistant features under complex channel conditions.
\item We design a VAE-based spoofed signal generation network with a multi-objective optimization strategy. This strategy directly guides spoofing performance at both the feature and classifier decision levels, thereby addressing the problem of high-concentration feature distributions and improving the overall attack effectiveness.
\item We build a simulation-based evaluation system incorporating typical channel variations (e.g., additive white Gaussian noise (AWGN), multipath fading, Doppler shift). Experimental results demonstrate that the proposed method maintains high spoofing performance in cross-channel scenarios and achieves over 95\% success rate under various channel parameters (e.g., SNR, K-factor).
\end{itemize}

The rest of this paper is organized as follows: Section~\ref{sec2} examines current attack strategies and identifies their issues; Section~\ref{sec3} introduces the relevant theoretical foundations; Section~\ref{sec4} details the proposed attack model and strategy; Section~\ref{sec5} details the framework of the proposed attack; Section~\ref{sec6} presents experimental results and analysis; finally, Section~\ref{sec7} concludes the paper.

\begin{table*}
\caption{Summary of attacks on RF fingerprint}
\centering
\footnotesize
\begin{adjustbox}{max width=\textwidth}
\begin{tblr}{
  colspec={
    Q[c,m,wd=0.7cm] 
    Q[c,m,wd=2cm]   
    Q[c,m,wd=2cm]   
    Q[c,m,wd=1cm]   
    Q[c,m,wd=1.5cm]   
    Q[c,m,wd=1cm] 
    Q[c,m,wd=1cm]   
    Q[c,m,wd=1.7cm]   
    Q[c,m,wd=1cm]   
    Q[c,m,wd=1cm]   
    Q[c,m,wd=1cm]   
    Q[c,m,wd=1cm]   
  },
  cells = {c},
  cell{1}{1} = {r=3}{},
  cell{1}{2} = {r=3}{},
  cell{1}{4} = {c=9}{},
  cell{2}{3} = {r=2}{},
  cell{2}{4} = {c=4}{},
  cell{2}{8} = {r=2}{},
  cell{2}{9} = {c=4}{},
  hline{1,4,12} = {-}{},
  hline{2} = {3-12}{},
  hline{3} = {4-7,9-12}{},
}
Ref & Attack Type & Target Feature & Attack assumptions &  &  &  &  &  &  &  & \\
 &  & Channel-resilient Feature & Known Legal Entity Information &  &  &  & {Consider \\attacker's RFF} & Channel conditions &  &  & \\
 &  &  & Model structure & Model parameters & RFF method & Fixed packet &  & AWGN & Multipath & Doppler & Non-Identical\\
\cite{ma2023white} & Evasion & \fullcirc & \fullcirc & \fullcirc & \fullcirc & \fullcirc & \emptycirc & \emptycirc & \emptycirc & \emptycirc & \emptycirc\\
\cite{liu2023robust} & Evasion & \emptycirc & \fullcirc & \fullcirc & \fullcirc & \emptycirc & \emptycirc & \fullcirc & \fullcirc & \fullcirc & \fullcirc\\
\cite{huang2023hidden} & Backdoor & \emptycirc & \fullcirc & \fullcirc & \fullcirc & \fullcirc & \emptycirc & \emptycirc & \emptycirc & \emptycirc & \emptycirc\\
\cite{danev2010attacks} & Impersonation & \emptycirc & \fullcirc & \fullcirc & \fullcirc & \fullcirc & \emptycirc & \emptycirc & \emptycirc & \emptycirc & \emptycirc\\
\cite{merchant2019securing} & Impersonation & \emptycirc & \fullcirc & \fullcirc & \fullcirc & \emptycirc & \emptycirc & \fullcirc & \emptycirc & \emptycirc & \emptycirc\\
\cite{karunaratne2021penetrating} & Impersonation & \emptycirc & \emptycirc & \emptycirc & \fullcirc & \fullcirc & \emptycirc & \fullcirc & \fullcirc & \emptycirc & \halfcirc\\
\cite{xu2022colluding} & Colluding Impersonation & \emptycirc & \fullcirc & \emptycirc & \fullcirc & \emptycirc & \fullcirc & \fullcirc & \halfcirc & \emptycirc & \emptycirc\\
\textbf{This Work} & \textbf{Colluding \\ Impersonation} & \fullcirc & \fullcirc & \fullcirc & \fullcirc & \fullcirc & \fullcirc & \fullcirc & \fullcirc & \fullcirc & \fullcirc
\end{tblr}
\end{adjustbox}
\begin{tablenotes}
\footnotesize
\item In this table, solid circles (\fullcirc[0.4em]) denote fully satisfied attack assumptions, hollow circles (\emptycirc[0.4em]) denote unsatisfied attack assumptions, and semi-solid circles (\halfcirc[0.4em]) denote partially satisfied attack assumptions, e.g., the similar channel condition required in~\cite{karunaratne2021penetrating}, the single-tap Rayleigh channel condition in~\cite{xu2022colluding}.
\end{tablenotes}
\label{tab:attack_summary}
\end{table*}

\section{Related Works}
\label{sec2}
This section reviews existing attacks on RFF systems and discusses their limitations. 
\subsection{Overview of existing attack methods}
Vulnerabilities in RFF systems are targeted within two main components: the neural networks used for identification and the extracted RFF features.

Neural network attacks can be executed in the following manner. Evasion attacks, for instance, create perturbation signals that overlay genuine transmissions, leading to misclassification. The algorithms for creating these perturbations often rely on gradient information or the neural network's decision boundaries. Ma et al.~\cite{ma2023white} reported success rates exceeding 90\% with the Fast Gradient Sign Method and Projected Gradient Descent on channel-independent spectrogram features. Liu et al.~\cite{liu2023robust} designed algorithms effective in both ideal and non-ideal channel scenarios, achieving similarly high success rates. 
Similarly, backdoor attacks incorporate corrupted samples into the training phase and employ trigger signals during validation to induce misclassification. Creating both these tainted samples and triggers necessitates comprehension of the neural network's decision mechanics. Huang et al.~\cite{huang2023hidden} attained nearly 100\% success in attacks, while preserving system accuracy on untainted samples, effectively ensuring stealth.

Impersonation attacks aim directly at RFF features by replicating or altering them, necessitating partial insight into the neural network's identifier. Danev et al.~\cite{danev2010attacks} investigated methods like signal and feature replay. Merchant and Nousain~\cite{merchant2019securing} utilized Generative Adversarial Networks (GANs) for creating spoofed signals and boosting defense. Karunaratne et al.~\cite{karunaratne2021penetrating} initiated black-box attacks with reinforcement learning using solely binary feedback. Xu et al.~\cite{xu2022colluding} employed colluders to overcome hardware limitations, achieving success rates exceeding 90\%.

\subsection{Limitations of current approaches}

While these works have made significant contributions, several limitations merit consideration:

\subsubsection{Information assumptions} Many studies assume white-box scenarios with comprehensive system knowledge. Although~\cite{xu2022colluding} reduced these requirements and~\cite{karunaratne2021penetrating} introduced black-box approaches, strong assumptions remain prevalent.

\subsubsection{Channel modeling} Several works employ simplified channel models, such as AWGN-only conditions or direct signal superposition without channel effects. While recent studies incorporate multipath fading, they often assume attackers experience similar channel conditions as legitimate users—an assumption rarely valid in practice.

\subsubsection{Attack targets} Current approaches primarily focus on I/Q-level mimicry, which becomes challenging under complex channels. The similarity in time-domain I/Q data may not translate to similarity in channel-robust features, particularly under significant channel variations.

Table~\ref{tab:attack_summary} provides an overview of attack types, target characteristics, and underlying assumptions. 
In summary, current RFF impersonation attack research often relies on simplified channel assumptions, limiting practical efficacy. This work addresses these limitations by introducing channel-resistant RFF into attack scenarios and proposing difference measurement at the feature and classifier decision levels to effectively target these robust features.

\section{Preliminary of Channel-Resistant RFF}
\label{sec3}
This section covers the system fundamentals and derives the channel-resistant RFF that remains reliable across multipath environments.
\subsection{System Model}
In an RFF-based authentication system, a transmitter modulates a baseband signal $s(t)$ and transmits it through its RF front-end. During this process, unique distortions arise due to hardware imperfections specific to the device, including power amplifier nonlinearity, local oscillator frequency offset, and I/Q imbalance, et al. These distortions resulting from hardware are expressed as a function $f(\cdot)$ applied to the baseband signal. The transmitted signal then propagates through a wireless channel with impulse response $h(t)$ and is corrupted by additive white Gaussian noise $n(t)$. The signal $r(t)$ received at the receiver is represented by
\begin{equation}
    r(t) = f(s(t)) * h(t) + n(t)
\end{equation}
where $*$ denotes convolution. Both hardware and channel distortions are present in $r(t)$. Since the fingerprint at the receiver’s front-end is identical for a given receiver, this model omits modeling the receiver’s fingerprint. The goal of RFF-based identification is to extract device-specific features that reflect the impact of $f(s(t))$, while being robust to variations in $h(t)$ and $n(t)$.

\subsection{Channel-Resistant RFF}
To isolate hardware-specific distortions from channel variations, channel-resistant RFF methods aim to extract features that depend only on hardware imperfections, regardless of the wireless channel conditions. Among these, CLPS features exhibit outstanding RFF-identification performance over unknown channel statistics~\cite{tang2024causal}.

CLPS begins by computing the power spectrum (PS) of the received signal, which is estimated as the squared magnitude of the Fast Fourier Transform (FFT) of the signal’s autocorrelation function. Subsequently, the logarithmic PS is obtained by taking the logarithm of the PS $P_r(\omega)$ of $r(t)$. This logarithmic scale representation can be expressed as:
\begin{align}
    P_{r}(\omega)_{\log} = &\log(\frac{1}{T}) + 2*\log|\text{FFT}(f(s(t)))| \notag
    \\&+ 2*\log|\text{FFT}(h(t))|
\end{align}
where $T$ denotes the duration of the captured signal.

 Given that the wireless channel experiences minimal variation over short durations, it can be reasonably assumed that the channel remains constant for each preamble. Based on this assumption, the CLPS is obtained by subtracting its mean from $P_{r,\log}(\omega)$ to mitigate channel variations:
\begin{align}
\text{CLPS}(\omega) =&P_{r}(\omega)_{\log} - \text{mean}(P_{r}(\omega)_{\log}) \notag
\\=&2*(\log|\text{FFT}(f(s(t))| \notag
\\&-mean(\log|\text{FFT}(f(s(t))|))
\end{align}

This formulation removes the channel-dependent term and preserves the spectral features introduced by hardware imperfections. Consequently, with a fixed baseband preamble, CLPS operates as a channel-resistant RFF that maintains stability, regardless of changing channel conditions~\cite{yang2021specific}.

\section{Impersonation attack strategy under complex channel conditions}
\label{sec4}
In this section, we will introduce the attack model and attack strategy, in order to clarify the attack scenario, assumptions, and the roles of each member within this scenario.

\subsection{Attack Model}

\begin{figure}[t]
\centering
\includegraphics[width=\linewidth]{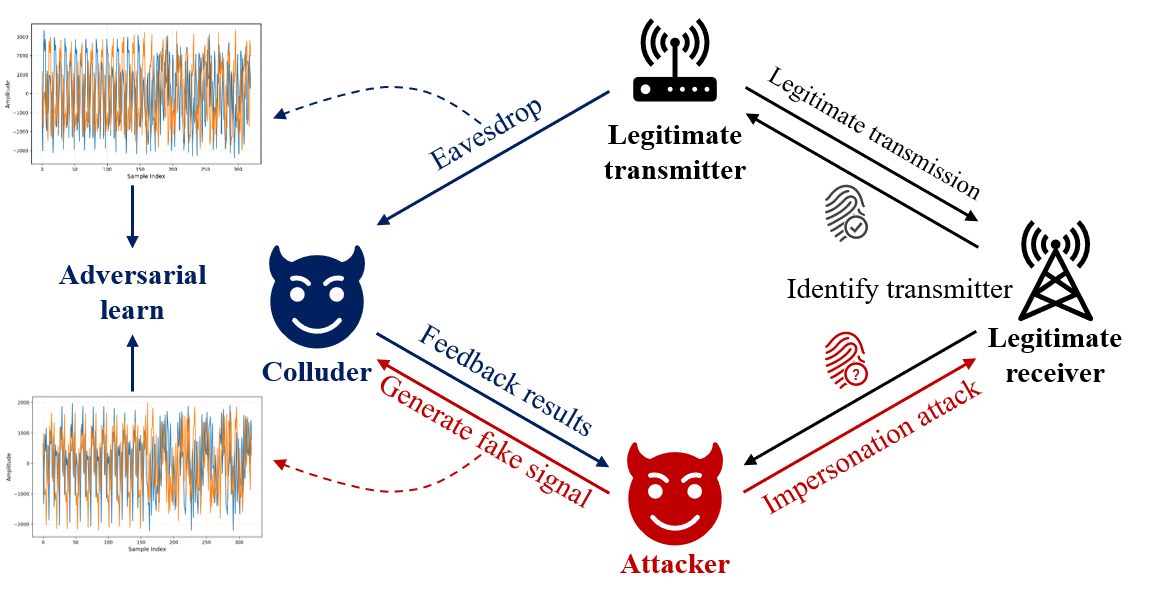}
\caption{The Proposed Collusion-Driven Impersonation Attack Model Illustration}
\label{impersonation attack}
\end{figure}

The attack model is shown in Fig.~\ref{impersonation attack}. We consider a collusion-driven impersonation attack targeting channel-resistant RFF systems. The scenario consists of the following four primary entities: a set of legitimate transmitters, a legitimate receiver, an attacker, and a colluder. Each transmitter periodically sends packets that include a fixed baseband preamble. Each transmitter introduces distinct distortions due to hardware imperfections, like I/Q imbalance, affecting the signal and acting as the device’s specific RFF.

During propagation, the signal experiences channel variations including multipath fading, Doppler shift, and AWGN. The legitimate receiver extracts a channel-resistant RFF, from the received signal's preamble segment and performs device identification via a deep neural network classifier. Here we use CLPS as an example, and the approach can later be extended to other channel-independent RFF features.

The attacker aims to impersonate a specific legitimate transmitter by generating spoofed signals that, when passed through its own transmitter hardware and an unknown wireless channel, yield CLPS features at the receiver that resemble those of the target device. In this model, the attacker does not possess knowledge of the receiver's channel conditions or the exact effect of its own hardware on the transmitted signal.

To overcome these limitations, the attacker collaborates with a colluder. The colluder is located at a different position and is capable of receiving transmissions from both the attacker and the target transmitter. It is assumed that the colluder has full knowledge of the CLPS feature extraction method and the classifier used by the legitimate receiver. Furthermore, the colluder employs the same RFF identification method as the legitimate receiver. This setup allows the attacker to train a spoofed signal generation network with limited prior information about the target environment. The use of CLPS ensures that even under different channels between the colluder and the receiver, consistent device features can be extracted.

The overall attack assumption settings are summarized in Table~\ref{tab:attack_summary}. The attacker is unaware of the impact of its own hardware impairments and relies on a colluding impersonation strategy to circumvent and optimize for this. To address the core challenge that attackers cannot obtain the channel conditions between devices, we resolve this issue by introducing channel-resistant RFF, aiming to enhance the feasibility of impersonation attacks in realistic channel environments. 
This study addresses an attack scenario operating under a white-box assumption, wherein the attacker has partial knowledge of the legitimate system, including classifier structures, parameters, known baseband signals, and fingerprint extraction techniques. The public access to various RFF methodologies provides a basis for this assumption. In practice, attackers might acquire details about the target RFF system, including feature extraction methods and protocol specifications, by consulting relevant technical documents or employing social engineering tactics. Furthermore, considering the widespread use of deep learning models, the attacker may also leverage existing model stealing techniques~\cite{oliynyk2023know}, in conjunction with eavesdropped legitimate signals, to train a surrogate model that approximates the behavior of the legitimate classifier. This surrogate model can then be used to guide the generation of spoofed signals during the attack process, without requiring knowledge of the exact classifier.

\subsection{Attack Strategy}
\begin{figure}[t]
\centering
\includegraphics[width=\linewidth]{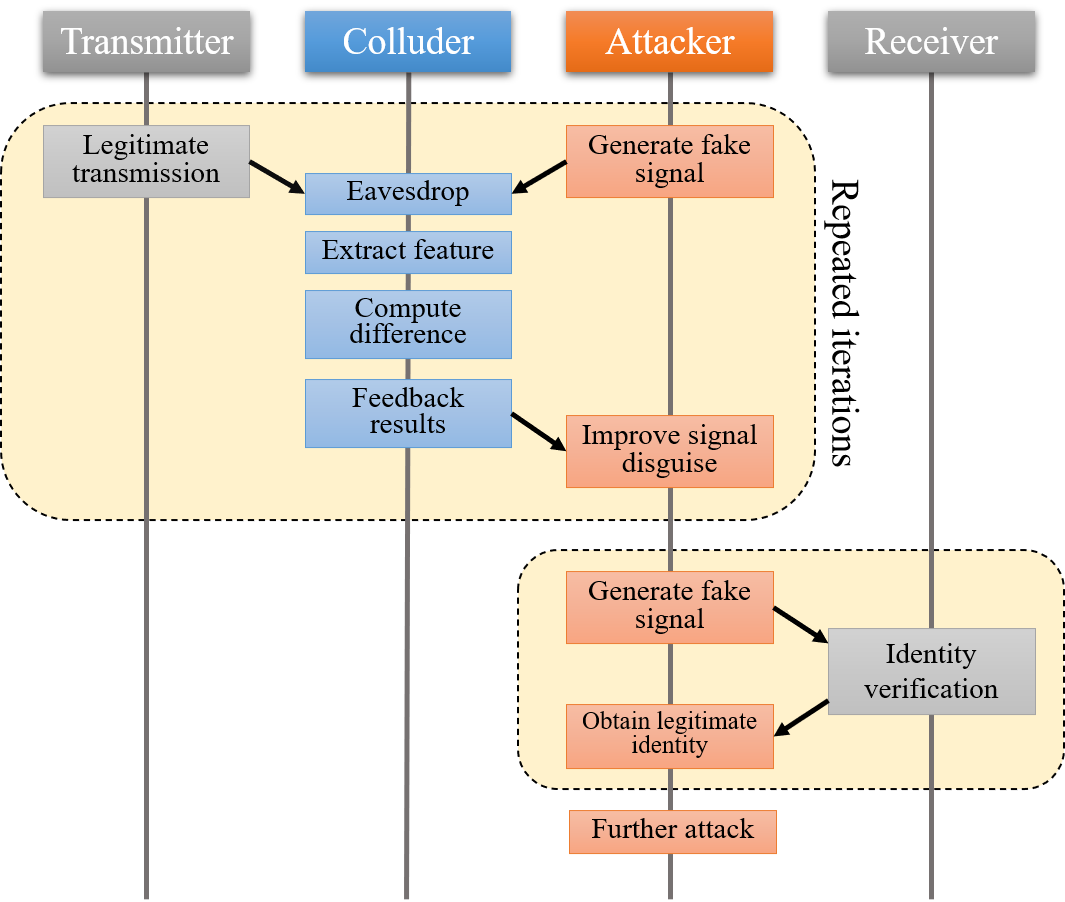}
\caption{The workflow of the proposed collusion-driven impersonation attack.}
\label{attack strategy}
\end{figure}

Fig.~\ref{attack strategy} illustrates the procedure of the suggested impersonation attack method. The attacker seeks to execute an impersonation attack where their transmitted signals are recognized as originating from the intended legitimate transmitter. Unlike traditional RF-level spoofing, which replicates hardware-induced distortions, the attacker aims to modify its signal to ensure that the spoofed signal exhibits a similar feature representation to that of the target signal. Accordingly, the attacker must learn the latent feature distribution associated with the target transmitter’s transmissions and then iteratively adapt its own signal to conform to this representation, thereby realizing the impersonation attack.

Considering the different channel conditions, the attacker cannot confirm the effect of its own hardware imperfections and channel variations; a colluder works with the attacker to help improve the way the attacker spoofs its RFF features. The colluder can be distant from the legitimate receiver. It captures the altered signal from the adversary and monitors the transmission from the target. The colluder then extracts features akin to the legitimate receiver and examines the differences. Given that the colluder utilizes the same classifier as the legitimate receiver, it can discern the receiver's classification outcome for the spoofed signal and subsequently send critical details regarding the inconsistencies to the attacker. With the feedback, the attacker can improve its signal disguise to minimize the feature distance between its own and the target's RFF representation. Through an iterative process, the attacker can gradually achieve a high RFF feature similarity to the target transmitter.

After finalizing the preparations for the attack, the attacker transmits optimized spoofed signals to the receiver. If the impersonation attack is successful, the receiver will misidentify the attacker's transmission as originating from the target transmitter and subsequently return the classification result to the attacker. Thereafter, the attacker can exploit the misclassifications to launch further attacks as a legitimate user.

\section{VAE-based Impersonation Attack Framework against Channel-Resistant RFF}
\label{sec5}

This section first presents the overall framework of the proposed impersonation attack and then derives the multi-objective loss function that drives the attack.

\subsection{Attack Implementation Framework}

\subsubsection{Spoofed Signal Generation Network}

The two mainstream generative models for wireless signal spoofing are GAN and VAE. The previous work of~\cite{xu2022colluding} adopted GAN, yet we opt for VAE. In a GAN, the generator depends on the discriminator's feedback. However, in RF fingerprints, channel robust features are so unique that the discriminator soon gains overconfidence, ceasing to offer valuable guidance and leading to stalled training. The VAE circumvents this by utilizing a defined loss function, combining reconstruction error with Kullback-Leibler (KL) regularization, which ensures a consistent gradient during training. Consequently, the VAE continuously aligns the spoofed signal with the target device's RF fingerprint, avoiding the premature convergence or training failure typical with GANs.

\begin{figure}[t]
\centering
\includegraphics[width=0.8\linewidth]{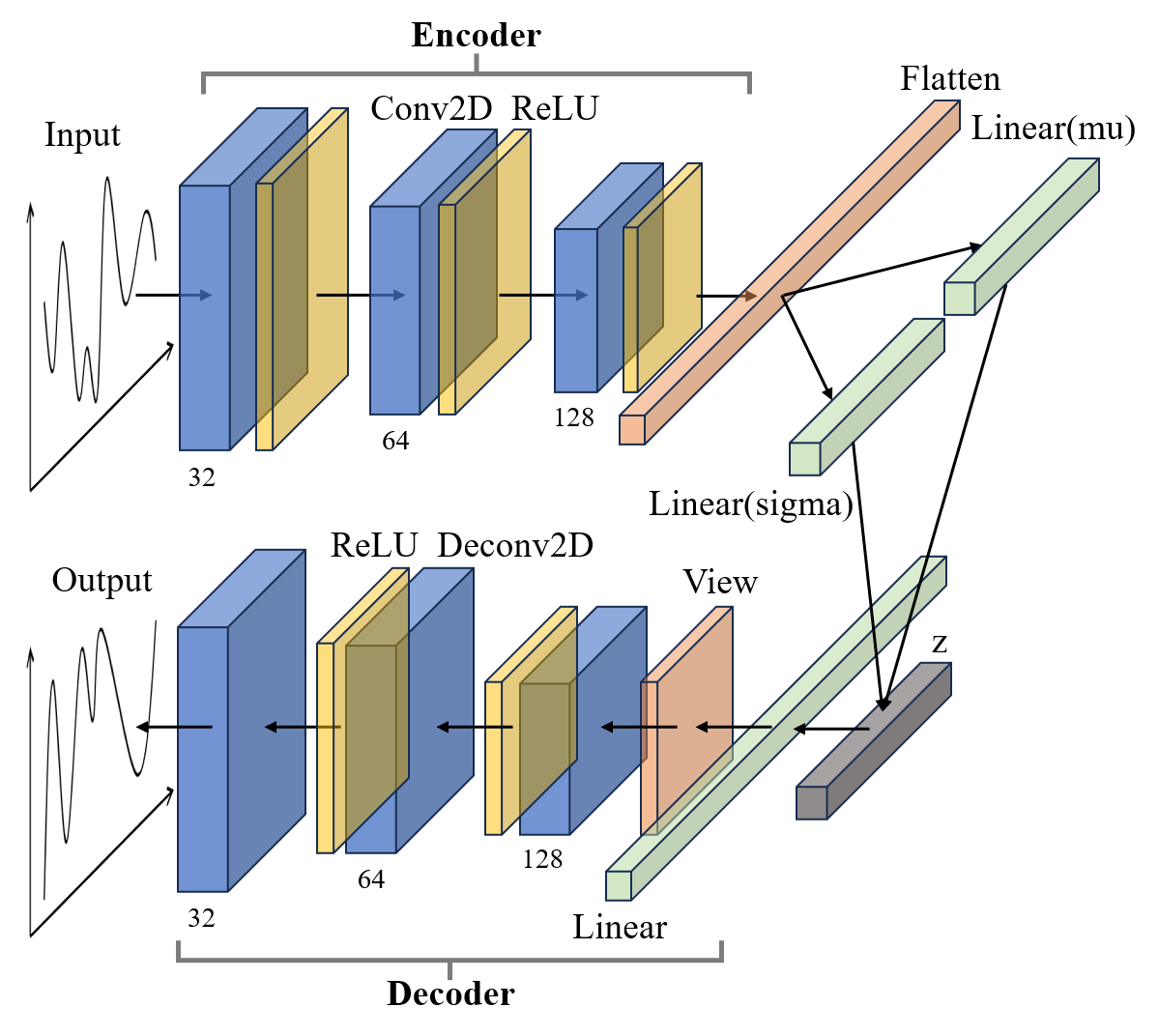}
\caption{Proposed spoofed signal generation network based on VAE model.}
\label{vaemodel}
\end{figure}

Therefore, as depicted in Fig. \ref{vaemodel}, we develop a spoofed signal generation network utilizing the VAE model. 
The proposed network is composed of an encoder, decoder, and a latent sampling module. The encoder includes three convolutional layers with (1×5) kernels, stride 2, and ReLU activations, which progressively compress the input into a compact latent space. Two fully connected layers produce the mean and log-variance of a 128-dimensional latent vector. The decoder operates as the inverse of the encoder, beginning with a fully connected layer and followed by three transposed convolutional layers to reconstruct the input's original shape.

\subsubsection{CLPS Feature Extraction and Classifier}

Algorithm~\ref{alg:clps} outlines the method for CLPS extraction, used by both the legitimate receiver and the colluder, with the goal of isolating channel-resistant RFF. Section~\ref{sec2} discusses how these CLPS features, derived from the signal's preamble, mainly differ due to the device's hardware imperfections rather than data content or channel variability. These robust features allow the colluder to evaluate feature discrepancies based on impersonated and genuine signals consistently with the legitimate receiver.
\begin{algorithm}[t]
\caption{Extract CLPS feature}
\label{alg:clps}
\begin{algorithmic}[1]
\REQUIRE Received signal $\mathbf{R}$
\ENSURE Channel-resistant RFF $\mathbf{CLPS}$

\STATE $\mathbf{ACF} \gets \text{Autocorrelation}(\mathbf{R})$
\STATE $\mathbf{P}_{\text{spec}} \gets |\mathcal{FFT}(\mathbf{ACF})|^2$
\STATE $\mathbf{P}_{\log} \gets \log_{10}(\mathbf{P}_{\text{spec}})$
\STATE $\mathbf{CLPS} \gets \mathbf{P}_{\log} - \text{Mean}(\mathbf{P}_{\log})$
\RETURN $\mathbf{CLPS}$
\end{algorithmic}
\end{algorithm} 

We utilize a conventional ResNet-based classifier for both the receiver and colluder, adhering to the architecture detailed in \cite{tang2024causal}.

\subsection{Multi-objective Loss functions}
To facilitate successful impersonation attacks via the VAE-based framework, we construct a multi-objective loss function that balances waveform naturalness, feature alignment, and adversarial deception. This aims to generate signals that preserve legitimate waveform structures, closely mimic the target device's CLPS features, and effectively trick the classifier. The detailed loss function is as follows:
\begin{align}
\label{loss function}
    Loss = &{\lambda}_{recon} * loss_{recon} + {\lambda}_{KL} *loss_{KL} \notag
    \\+ &{\lambda}_{CLPS}*loss_{CLPS}+{\lambda}_{Cls}*loss_{Cls}
\end{align}
where ${\lambda}$ are hyperparameters for balancing each loss term. In our experiment, they are set to 2.0, 0.1, 1.0, and 0.5, respectively.

\subsubsection{Reconstruction Loss ($loss_{recon}$)}
Measures the mean squared error (MSE) between the altered baseband signals and the initial input signals. This component guarantees that the spoofed signals preserve appropriate temporal characteristics, thereby avoiding the generation of aberrant samples that could arise when optimizing solely based on other loss functions. The loss is calculated as:
\begin{align}
\label{recon loss}
    loss_{recon} = ||r(t)_{recon} - r(t)_{input}||_2^2 \tag{4a}
\end{align}

\subsubsection{KL Divergence Loss ($loss_{KL}$)}
Regularizes the latent space distribution by minimizing the KL divergence between the learned distribution and a standard normal distribution. This term primarily serves to maintain training stability rather than strictly enforcing output samples to follow a particular distribution. The loss is calculated as:
\begin{align}
\label{kl loss}
    loss_{KL} = -0.5\sum(1 + \log(\sigma^2) - \mu^2 - \sigma^2) \tag{4b}
\end{align}

\subsubsection{CLPS Loss ($loss_{CLPS}$)}
Computes the mean squared error between the CLPS features of reconstructed signals and target CLPS features. This term ensures the spoofed signals' CLPS closely match those of the target device. The loss is calculated as:
\begin{align}
\label{clps loss}
    loss_{CLPS} = ||CLPS_{{recon}} - CLPS_{{target}}||_2^2 \tag{4c}
\end{align}

\subsubsection{Classification Loss ($loss_{Cls}$)}
Uses cross-entropy loss to measure how effectively the spoofed signals deceive the target classifier into predicting the desired device label. This adversarial term drives the VAE to produce signals that successfully fool the final classification decision. The loss is calculated as:
\begin{align}
\label{cls loss}
    l&oss_{Cls} = CrossEntropy(y_{recon},y_{target}) \tag{4d}
\end{align}
With the multi-objective loss fully specified, the design of the impersonation attack is now complete. 
\section{Performance Evaluation}
\label{sec6}
This section details the experimental setup and presents the results alongside comparative evaluations.
\subsection{Simulation Setup}

The simulation settings are summarized in Table \ref{setup}. The specific explanation of the simulation setup is as follows.

\begin{table}[t]
    \renewcommand\arraystretch{1.2}
    \renewcommand{\tabularxcolumn}[1]{m{#1}}
    \centering
    \caption{Summary of Simulation Setup}
    \label{setup}
    \begin{tabularx}{0.48\textwidth}{|l|X|}
    \hline
        \textbf{Component} & \textbf{Setting / Value} \\ \hline
        Devices & 10 legitimate + 1 attacker \\ \hline
        I/Q Imbalance & Fixed per device, gain $\in$ [-0.3, 0.3], phase $\in$ [-15°, 15°] \\ \hline
        Channel Models & AWGN, Rician, Rayleigh \\ \hline
        Sample Length & 5120 complex I/Q samples \\ \hline
        Dataset Size & 55000 training samples + 5500 test samples \\ \hline
    \end{tabularx}
\end{table}

\begin{table}[t]
    \renewcommand\arraystretch{1.2}
    \renewcommand{\tabularxcolumn}[1]{m{#1}}
    \centering
    \caption{Default Training Parameters}
    \begin{tabular}{|l|l|}
    \hline
        \textbf{Parameter} & \textbf{Setting / Value} \\ \hline
        Channel Type & Rician + AWGN \\ \hline
        Delay Vector (ns) & [0, 50, 110, 170, 290, 310] \\ \hline
        Gain Vector (dB) & [0, -3, -10, -18, -26, -32] \\ \hline
        K Factor & 5 \\ \hline
        Max Doppler Shift (Hz) & 10 \\ \hline
        $E_b/N_0$ (dB) & 10 \\ \hline
    \end{tabular}
    \label{trainpara}
\end{table}

\subsubsection{Device Configuration}
The simulation involves 10 legitimate transmitters and a single attacker, each characterized by distinct I/Q imbalance values to simulate hardware-specific distortions. Every device employs a fixed baseband preamble per IEEE 802.15.4 standards and transmits 5120-sample signal segments. The gain and phase imbalance parameters, based on realistic hardware constraints, align with previous research~\cite{xu2022colluding}.

\subsubsection{Channel Configuration}
We examine robustness across various propagation scenarios by simulating four typical environments: AWGN, Indoor Office Channel A/B, and Vehicular Channel A, utilizing ITU-R M.1225 models. Channels are synthesized using MATLAB's \texttt{comm.AWGNChannel} and \texttt{comm.RicianChannel} functions, featuring independent pathways for each entity. The default training configuration employs Rician fading with a $K$ factor of 5, a six-tap multipath profile, and $E_b/N_0 = 10$ dB, as outlined in Table~\ref{trainpara}.

\subsubsection{Dataset Description}
The dataset is composed of simulatively generated signals. Each device produces 5000 training segments and 500 testing segments, each with 5120 I/Q samples. CLPS features are extracted from each segment to train a ResNet-based classifier. To generate spoofed signals, the attacker feeds perturbed baseband preambles into its VAE model. The colluder supplies the target CLPS features and classifier outputs required for loss calculation during training.
\subsection{Effectiveness of the Proposed Impersonation Attack}

\begin{figure}[t]
    \centering
    \subfigure[]{
    \includegraphics[width=0.48\linewidth]{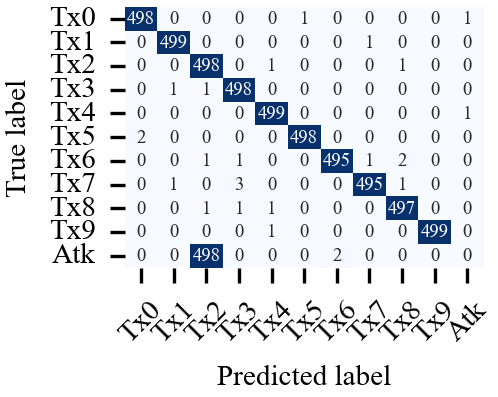}
    \label{asr2}
    }
    \hspace{-5mm}
    \subfigure[]{
    \includegraphics[width=0.48\linewidth]{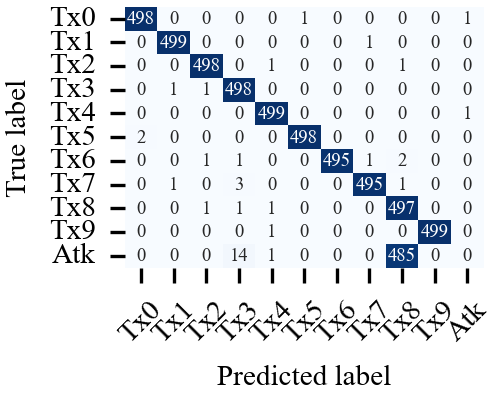}
    \label{asr8}
    }
    \quad
    \caption{Classification result under the proposed collusion-driven impersonation attack. (a) Target on legitimate transmitter 2. (b) Target on legitimate transmitter 8.}
    \label{asr_device}
\end{figure}

We first assess the proposed collusion-driven 
 attack's efficacy by analyzing classification results from the legitimate classifier, as illustrated in Fig.~\ref{asr_device}. For visualization purposes, Tx 2 and 8 were chosen at random. Fig.~\ref{asr_device} shows that the legitimate classifier accurately identifies legitimate devices, while spoofed signals aimed at specific devices are classified as the intended target. This implies that the attacker's spoofed signals possess the RFF characteristics of the target device, leading to a high likelihood of misclassification across various devices, indicating a probable attack success.

\subsection{Comparison with Existing Attack Strategies}

\begin{figure}[t]
    \centering
    \subfigure[]{
    \includegraphics[width=0.48\linewidth]{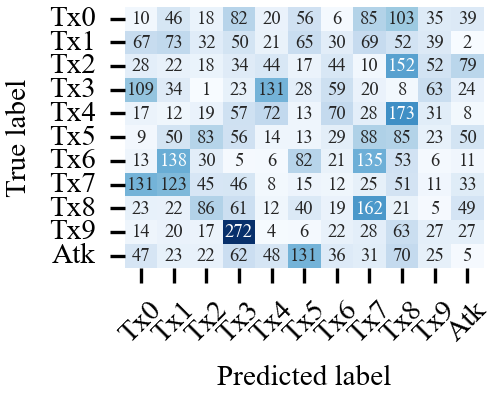}
    \label{figwoCLPS}
    }
    \hspace{-5mm}
    \subfigure[]{
    \includegraphics[width=0.48\linewidth]{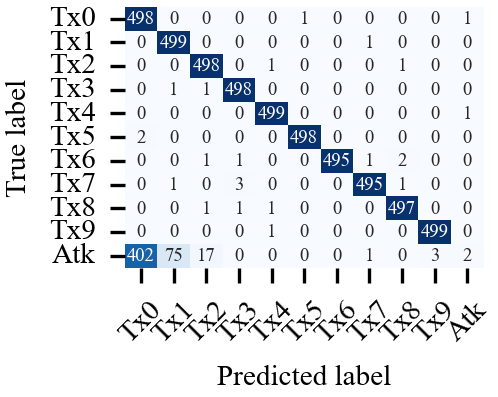}
    \label{figwoVAE}
    }
    \caption{Classification result of the existing attack strategy~\cite{xu2022colluding}, target on Tx2. (a) Classification result when targeting I/Q data. (b) Classification result when using GAN.}
\end{figure}

We evaluate our approach in relation to current attack methods, specifically where attackers alter raw I/Q data via a GAN-based network~\cite{xu2022colluding}. Experiments involve varying attack targets (I/Q data and CLPS) and generation networks (GAN and VAE), validating the effectiveness of our attack framework.

\begin{figure}[t]
    \centering
    \includegraphics[width=0.8\linewidth]{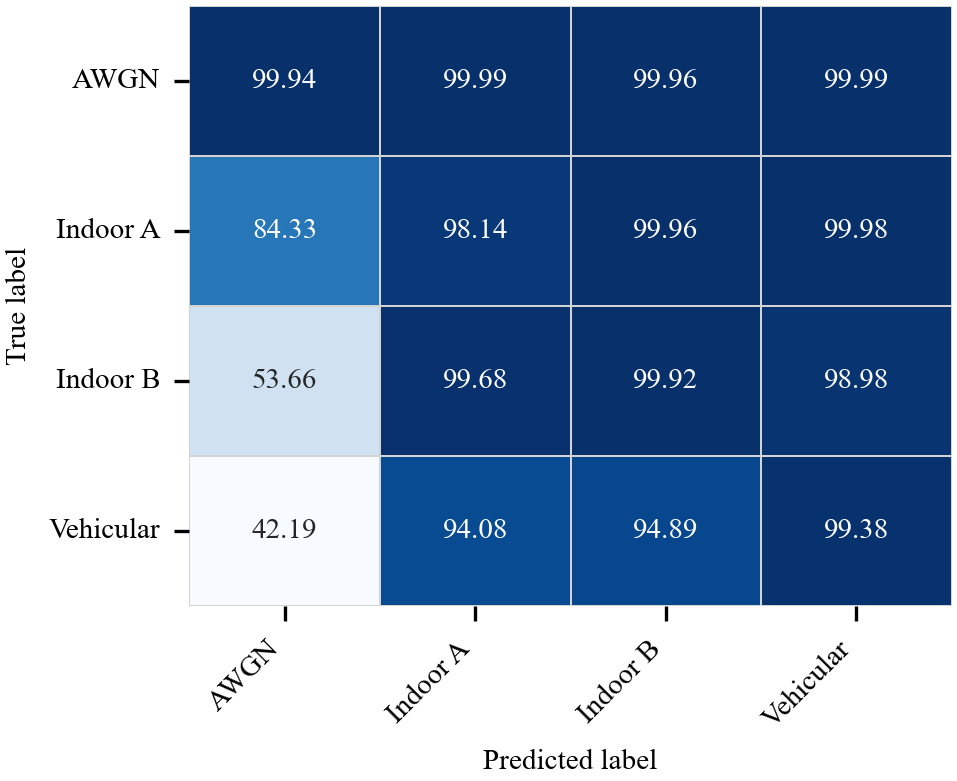}
    \caption{Cross-channel mean attack success rate.}
    \label{cross}
\end{figure}

\begin{figure*}[t]
    \centering
    \includegraphics[width=0.7\textwidth]{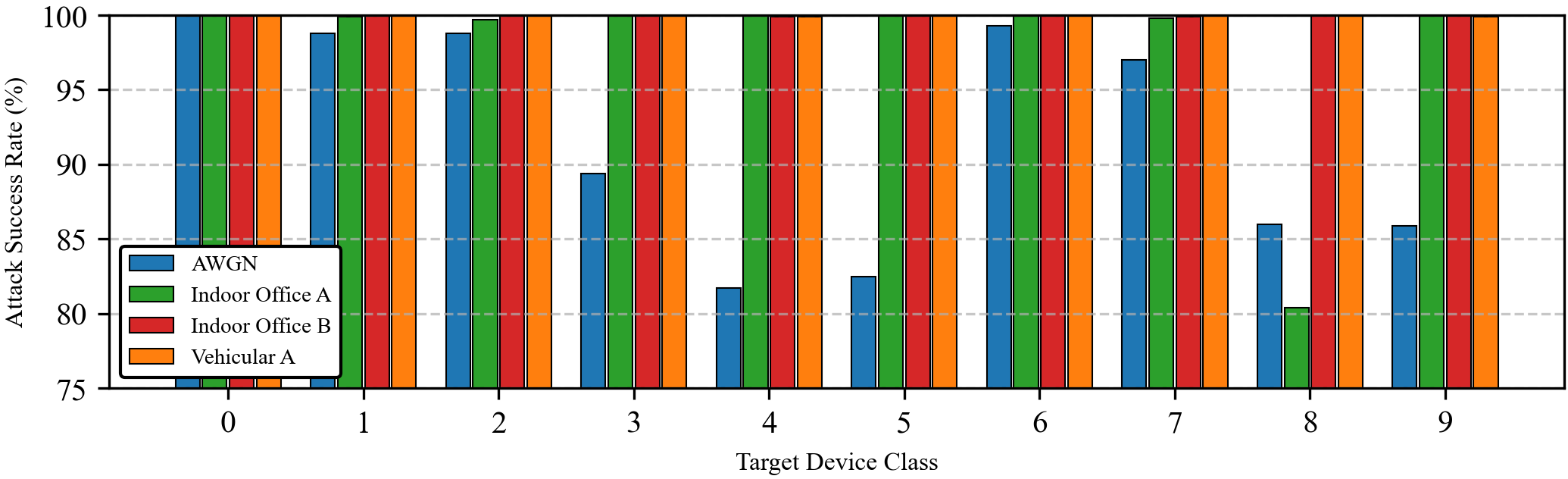}
    \caption{Attack success rates across target devices under models trained under different channels.}
    \label{all_asr}
\end{figure*}

\subsubsection{I/Q-Domain Attacks}

To evaluate the effectiveness of channel-resistant RFF modeling, we conduct a baseline experiment using raw I/Q samples as the input feature, without the CLPS extraction stage. Both the receiver and the colluder directly operate on I/Q data, and the spoofed signal is generated by a VAE trained to mimic the I/Q characteristics of the target transmitter. As shown in Fig.~\ref{figwoCLPS}, this approach fails to generate effective spoofed signals, resulting in almost random classification at the receiver.

The failure arises because I/Q data is extremely susceptible to channel variations. Lacking channel-resistant RFF, colluders obtain unstable features, which hinders them from offering reliable feedback for spoofed signal generation. This instability obstructs model convergence, making I/Q-level impersonation impractical in realistic multipath environments. It underscores the need for channel-resistant features such as CLPS for effective attack implementation.

\subsubsection{GAN-Based Signal Generation}

To further assess the attack design, we compare the proposed VAE-based generation method with the GAN-based approach. Both models aim to generate spoofed signals whose CLPS features match the target. As shown in Fig.~\ref{figwoVAE}, the GAN model fails to achieve effective impersonation, with spoofed signals consistently misclassified as a different device (e.g., Tx0 instead of the intended Tx2).

The GAN's performance declines due to high similarity in CLPS samples, causing fast discriminator convergence and resulting in vanishing gradients for the generator, thereby obstructing training. Conversely, VAE training uses explicit reconstruction and feature-aligned losses for consistent network updates, allowing for a smooth latent space and improved approximation of the target device's CLPS distribution. Thus, the proposed VAE framework improves spoofing stability and effectiveness, proving superior in generating high-fidelity RFF features for impersonation in channel-resistant modeling.

\subsection{Attack Success Rate in Complex Channel Conditions}
The proposed attack's effectiveness is assessed under intricate channel conditions. The attack success rate (ASR), defined as the proportion of signals mistakenly identified as legitimate targets to those transmitted by the attacker, is calculated. Evaluation begins with varied channel types and parameters.

\subsubsection{Performance under Different Channel Types}

We trained attack models for each of the four channel conditions and conducted comprehensive cross-channel evaluations. The results in Fig.~\ref{cross} indicate that, except for the model trained under the AWGN-Only channel, the other attack models demonstrate high resilience across different channel conditions. Notably, the model trained under the Vehicular channel condition achieves the best overall performance. In contrast, models trained under the Indoor channel condition show a slight drop in ASR when tested on the Vehicular channel. This is because the Vehicular channel presents a more complex multipath environment than the Indoor channel. As a result, a model trained under Vehicular conditions can better adapt to more challenging channel scenarios, thereby achieving superior performance. These results also reflect the impact of mismatched channel conditions between the colluder and the receiver, indicating that the impersonation attack performance degrades as their channel divergence increases.

To further understand the impact of channel mismatch and device-specific factors on spoofing effectiveness, Fig.~\ref{all_asr} presents a detailed breakdown of attack success rates across different devices, where the test channel is fixed to Indoor Office Channel A. The results show a consistent trend with Fig.~\ref{cross}, in that models trained under all channel conditions except AWGN maintain high ASR across most target devices—often nearing 100\%. Particularly, models trained under Vehicular and Indoor Office Channel B conditions continue to achieve strong performance, consistent with their ability to generalize better in the presence of multipath fading and other channel variations.

However, a notable performance drop is observed for attack models trained under AWGN and Indoor Office Channel A when evaluated across different devices, with the ASR decreasing to around 80\% in certain cases. This drop is primarily due to increased signal distortion under mismatched channel conditions, which reduces the similarity between the crafted spoofed signal and the target signal. Additionally, in the simulation setup, hardware impairments are limited to modeling IQ imbalance only. As a result, devices with similar IQ imbalance parameters may become less distinguishable, making them harder to spoof or more susceptible to misclassification, thereby reducing the attack success rate for those specific model-device pairs. This highlights the need to consider a wider range of hardware features in modeling device-specific characteristics to further improve spoofing robustness.

\subsubsection{Impact of $E_b/N_0$}

\begin{figure}[t]
    \centering
    \subfigure[]{
    \includegraphics[width=0.48\linewidth]{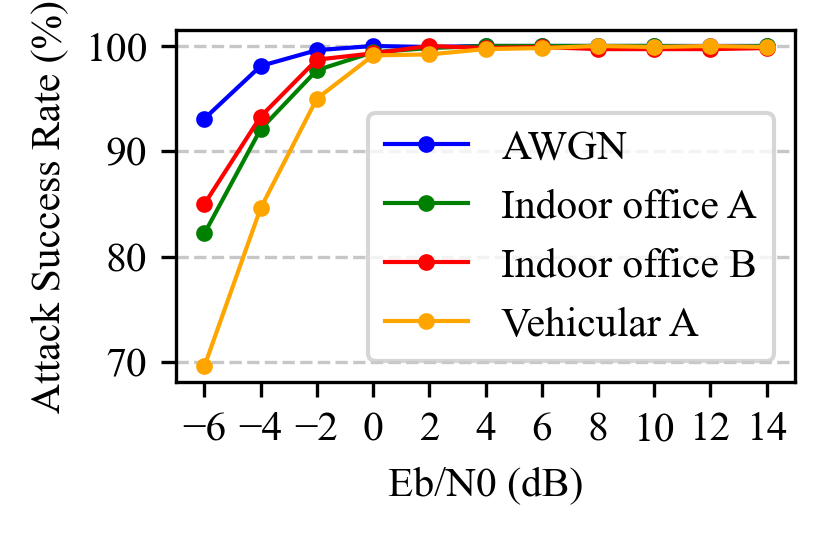}
    \label{ebn0}
    }
    \hspace{-5mm}
    \subfigure[]{
    \includegraphics[width=0.48\linewidth]{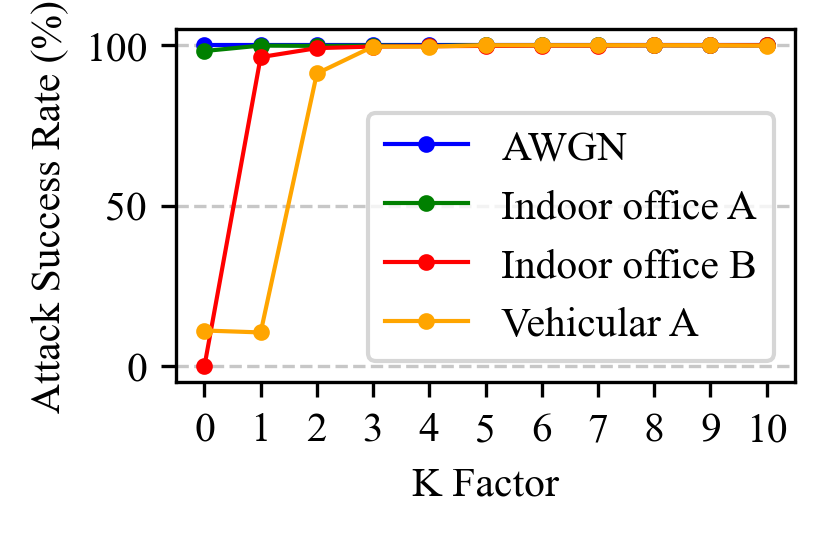}
    \label{kfactor}
    }
    \caption{ASR performance under different channel parameters. (a) ASR vs $E_b/N_0$ under different channels. (b) ASR vs K-factor under different channels.}
    \label{asr}
\end{figure}

We evaluate the attack performance in different SNR, the result is shown in Fig. \ref{ebn0}. From the figure, it can be observed that the success rate of impersonation attacks increases with the rise of $E_b/N_0$. This is in line with expectations, as a higher $E_b/N_0$ indicates that the legitimate receiver experiences less distortion in the received spoofed signals. 

Furthermore, it can be observed that channels A and B maintain a generally consistent attack rate under different $E_b/N_0$ conditions. Although the vehicular channel performs relatively poorly at lower $E_b/N_0$ levels, the attack rate under this channel condition quickly approaches 100\% as $E_b/N_0$ increases. This indicates that the proposed attack has relatively weak channel robustness at low $E_b/N_0$ levels, but exhibits good channel robustness as $E_b/N_0$ rises (for $E_b/N_0$$\geq$0).

\subsubsection{Impact of K-factor}

To validate the robustness of the proposed attack, we assess its performance over different K-Factors (e.g., K = [0, 1, ..., 10]). The results are presented in Fig. \ref{kfactor}. This figure shows that the attack success rate (ASR) in the Indoor Office Channel A is stable and comparable to that in an AWGN channel across varying K-factors. Notably, the ASR is approximately 97.40\% at K = 0 and rises to nearly 100.0\% as K grows to 10. Using MATLAB’s comm.RicianChannel function, a K-factor of zero simulates a Rayleigh fading model. The findings highlight the attack's robustness to variations in the K-factor, ensuring steady performance under line-of-sight (LoS, K\textgreater0) and non-line-of-sight (NLoS, K = 0) conditions, assuming consistent channel characteristics.

Conversely, in Indoor Office Channel B and Vehicular Channel A, attack effectiveness declines markedly at low K-factors. For example, in Indoor Office Channel B, the ASR is merely 0.20\% at K=0, but climbs to 100.0\% with improved channel conditions. Similarly, in Vehicular Channel A, the ASR reduces to 12.40\% at K=0, escalating to 99.80\% at K=10. This is due to the lack of a direct LoS path in NLoS scenarios, where signal propagation depends fully on multipath reflections, causing significant fading and distortion. The steep degradation in NLoS conditions underscores the sensitivity of these channels to specific multipath variables. However, under LoS conditions, the proposed attack remains resilient to changes in the K-factor.

\section{Conclusion}
\label{sec7}
In this paper, we present a collusion-driven impersonation attack targeting channel-resistant RFF systems. By incorporating CLPS as a channel-invariant feature and introducing a VAE-based spoofed signal generation network, the attacker is able to mimic the target device’s RF fingerprint under complex wireless channel conditions. Leveraging a colluder that shares the same feature extractor and classifier as the legitimate receiver, the attacker iteratively optimizes signal generation to minimize discrepancies in both feature space and classification outcome. Extensive simulations across diverse scenarios—including AWGN, multipath fading, and Doppler shift—demonstrate that the proposed attack achieves stable success rates (\textgreater95\%) across fading channels and SNR conditions. These results reveal the practical feasibility and significant threat posed by collusion-driven impersonation attack in realistic RFF systems with channel-resilient features.

Future work may explore adversarial training techniques to improve the classifier’s robustness against synthesized spoofed samples. Additionally, active defense strategies at the legitimate transmitter, such as embedding recoverable perturbations known only to the intended receiver, may offer effective protection by limiting the attacker’s ability to accurately observe and replicate legitimate transmissions.

\section*{ACKNOWLEDGMENT}
This work was supported in part by the Frontier Technologies Research and Development Program of Jiangsu under Grant BF2024065; in part by the National Natural Science Foundation of China under Grant 62171121 and Grant U22A2001;  in part by the Natural Science Foundation on Frontier Leading Technology Basic Research Project of Jiangsu under Grant BK20222001.

\bibliographystyle{unsrt}
\bibliography{ref.bib}

\end{document}